\documentclass[letterpaper, 10 pt, conference]{ieeeconf}  

\IEEEoverridecommandlockouts                              
\newcommand{\sofie}[1]{\textcolor{red}{#1}}

\title{\LARGE \bf
	Verifying the Unknown: Correct-by-Design Control Synthesis for Networks of Stochastic Uncertain Systems
}

\author{Oliver Sch\"{o}n$^{1}$, Birgit van Huijgevoort$^{2}$, Sofie Haesaert$^{2}$, and Sadegh Soudjani$^{1}$
		\thanks{This work is supported by the following grants: EPSRC EP/V043676/1, EIC 101070802, ERC 101089047, and NWO Veni 18244.
		}
		\thanks{$^{1}$Oliver Sch\"{o}n and Sadegh Soudjani are with the School of Computing, Newcastle University, Newcastle, NE4 5TG, UK.
			{\tt\small o.schoen2@ncl.ac.uk,sadegh.soudjani@ncl.ac.uk}}%
		\thanks{$^{2}$Birgit van Huijgevoort and Sofie Haesaert are with the Electrical Engineering Department, TU Eindhoven, The Netherlands
			{\tt\small b.c.v.huijgevoort@tue.nl, s.haesaert@tue.nl}}%
	}

\usepackage{longtable}
\usepackage{mathtools}
\usepackage{amsfonts,amsmath,amssymb}

\usepackage{amsthm}
\makeatletter
\renewcommand*\env@matrix[1][*\c@MaxMatrixCols c]{%
	\hskip -\arraycolsep
	\let\@ifnextchar\new@ifnextchar
	\array{#1}}
\makeatother
\usepackage{booktabs}
\usepackage{mathrsfs}  
\usepackage{tikz}
\usepackage{pgfplots}
\usetikzlibrary{positioning,calc,fit,arrows}
\usepackage{tkz-fct}
\usepackage{calc}
\usetikzlibrary{intersections}
\usetikzlibrary{decorations.shapes,shapes.geometric,shadows,arrows,automata,positioning,calendar,mindmap,backgrounds,scopes,chains,er,patterns,pgfplots.groupplots}
\tikzset{initial text={}, 
	double distance=2pt, 
	every state/.style = {draw = black, fill = grayfilling} 
}
\usetikzlibrary{calc}
\usepackage[load-configurations=version-1]{siunitx} 
\usepackage{tikz-3dplot}
\usepackage{subfig}
\usepackage{url}
\usepackage[author={Oliver}]{pdfcomment}
\pdfcommentsetup{color=red}
\usepackage[framemethod=TikZ]{mdframed}
\usepackage{algcompatible}
\usepackage{algorithm}

\makeatletter
\renewcommand{\fnum@algorithm}{\fname@algorithm{} \thealgorithm:}
\makeatother

\usepackage[outdir=./]{epstopdf}

\newtheoremstyle{theoremdd}
{\topsep}
{\topsep}
{\itshape}
{0pt}
{\bfseries}
{:}
{ }
{\thmname{#1}\thmnumber{ #2}\boldmath\textbf{\thmnote{ (#3)}}} 
\theoremstyle{theoremdd}

\usepackage{xcolor}
\definecolor{lightblue}{rgb}{0.67, 0.9, 0.93}
\definecolor{lightgreen}{rgb}{0.67, 0.88, 0.69}
\definecolor{lightpink}{rgb}{1.0, 0.72, 0.77}
\definecolor{lightpurple}{rgb}{0.96, 0.73, 1.0}
\definecolor{lightyellow}{rgb}{0.98, 0.93, 0.37}
\definecolor{grayfilling}{gray}{0.95} 
\definecolor{grayshadow}{gray}{0.5} 
\colorlet{cherryred}{red!80!black}

\newtheorem{problem}{Problem}
\newtheorem{definition}{Definition}
\newtheorem{theorem}{Theorem}

\newenvironment{prob}
{\begin{mdframed}[backgroundcolor=grayfilling, shadow=true, shadowsize=5.5pt, shadowcolor=grayshadow]
		\begin{problem}}
		{\end{problem}\vspace{.4em}\end{mdframed}}
\newtheorem{assumption}{Assumption}       

\newtheorem{lemma}{Lemma}
\newtheorem{example}{Example}

\newcommand{\R}{\mathcal{R}} 
\newcommand{\Rx}[1]{\mathcal{R}_{x^{#1}}} 
\newcommand{\Ru}[1]{\mathcal{R}_{u^{#1}}} 

\newcommand{\norm}[1]{\left\lVert#1\right\rVert} 
\newcommand{\abs}[1]{\left| #1\right|} 
\newcommand{\cdf}[1]{\mathrm{cdf}\left( #1 \right)}
\newcommand{\offset}{\gamma}

\newcommand{\T}{^\top}
\newcommand{\dxp}{dx_+}
\newcommand{\dxhp}{d\hat x_+}
\newcommand{\xp}{x_+}
\newcommand{\xhp}{\hat x_+}

\ifdefined\U
\renewcommand{\U}{\mathbb{U}}
\else
\newcommand{\U}{\mathbb{U}}
\fi

\newcommand{\Uh}{\hat{\mathbb{U}}}
\newcommand{\given}{\;|\;}

\newcommand{\Interc}[1]{\mathscr{I}(#1)}

\newcommand{\satisfies}{\vDash}

\newcommand{\Tr}{\mathbf{t}}
\newcommand{\Trh}{\hat{\mathbf{t}}}

\newcommand{\X}{\mathbb{X}}
\newcommand{\Xh}{\hat{\mathbb{X}}}
\newcommand{\x}[1]{{x}_{#1}}

\newcommand{\M}{\mathbf M}
\newcommand{\Mh}{\widehat{\mathbf M}}
\newcommand{\Mt}{\widetilde{\mathbf M}}
\newcommand{\A}{\mathbb{U}}
\newcommand{\Ah}{\hat{\mathbb{U}}}

\newcommand{\ac}[1]{u_{#1}}

\newcommand{\rel}{\mathcal{R}}

\newcommand{\Hist}{\mathbb{H}}

\newcommand{\pol}{{\varphi}}

\newcommand{\CA}[1]{\mathsf{#1}}

\newcommand{\AP}{\mathsf{AP}}

\newcommand{\notltl}{\neg}
\newcommand{\andltl}{\wedge}
\newcommand{\orltl}{\vee}

\newcommand{\Next}{\ensuremath{\bigcirc}}

\newcommand{\Event}{\ensuremath{\ \diamondsuit\ }}
\newcommand{\Until}{\ \CA{U}\ }

\newcommand{\alphabeth}{\Sigma}
\newcommand{\word}{\boldsymbol{\pi}}

\newcommand{\letter}{l}
\newcommand{\True}{\operatorname{\mathsf{true}}}

\newcommand{\N}{\mathcal{N}} 
\newcommand{\GMM}{\N\hspace{-.7em}\N} 

\newcommand{\meas}{\nu}     
\newcommand{\po}{p}     
\newcommand{\pok}{\mathbf{p}}     
\newcommand{\pk}[1]{\pok\left(#1\right)}     
\newcommand{\borel}[1]{\mathcal{B}\left(#1\right)}
\newcommand{\eps}{\varepsilon}
\newcommand{\InF}{\mathbf{i}} 

\newcommand{\Ca}{{\mathbf{C}}}
\newcommand{\Cah}{\widehat{\mathbf{C}}}
\newcommand{\Y}{\mathbb{Y}}

\newcommand{\dist}{\mathbf{d}_{\Y}}

\newcommand{\W}{v}
\newcommand{\Wt}{\boldsymbol{\W}}
\newcommand{\fWt}{\bar{\Wt}} 
\newcommand{\sWt}{{\Wt}} 

\newcommand{\xh}[1]{\hat{x}_{#1}}

\newcommand{\uh}[1]{\hat{u}_{#1}}

\newcommand{\ach}[1]{\hat{u}_{#1}}

\newcommand{\cdotx}{\,\cdot\,}

\renewcommand{\P}{\mathbb{P}}

\usepackage{tcolorbox}
\usepackage{mathtools}

\newtcbox{\blueb}{nobeforeafter,tcbox raise base,boxrule=0.4pt,top=0mm,bottom=0mm,
	right=0mm,left=0mm,arc=1pt,boxsep=2pt,before upper={\vphantom{dlg}},
	colframe=blue!50!black,coltext=black!25!black,colback=blue!10!white}
\newtcbox{\redb}{nobeforeafter,tcbox raise base,boxrule=0.4pt,top=0mm,bottom=0mm,
	right=0mm,left=0mm,arc=1pt,boxsep=2pt,before upper={\vphantom{dlg}},
	colframe=red!50!black,coltext=black!25!black,colback=red!10!white}

\newtcbox{\bluebs}{nobeforeafter,tcbox raise base,boxrule=0.4pt,top=0mm,bottom=0mm,
	right=0mm,left=0mm,arc=1pt,boxsep=.5pt,before upper={\vphantom{dlg}},
	colframe=blue!50!black,coltext=black!25!black,colback=blue!10!white}
\newtcbox{\redbs}{nobeforeafter,tcbox raise base,boxrule=0.4pt,top=0mm,bottom=0mm,
	right=0mm,left=0mm,arc=1pt,boxsep=.5pt,before upper={\vphantom{dlg}},
	colframe=red!50!black,coltext=black!25!black,colback=red!10!white}

\robustify{\bluebs}
\robustify{\redbs}
\robustify{\blueb}
\robustify{\redb}

\usepackage{hyperref}
\hypersetup{
	colorlinks   = true,    
	urlcolor     = cyan,    
	linkcolor    = cyan,    
	citecolor    = cyan      
}

\newcommand{\old}[1]{{\color{gray} #1}}
\newcommand{\Oliver}[1]{{\color{red!80!black} [Oliver]: #1}}

\usepackage{tikz}

\graphicspath{{./Figures/}} 
\usepackage{blindtext}
\usepackage{float}
\usepackage{dblfloatfix}

\begin{document}
\maketitle
\thispagestyle{empty}
\pagestyle{empty}

\begin{abstract}
In this paper, we present an approach for designing correct-by-design controllers for cyber-physical systems composed of multiple dynamically interconnected uncertain systems. We consider networked discrete-time uncertain nonlinear systems with additive stochastic noise and model parametric uncertainty. Such settings arise when multiple systems interact in an uncertain environment and only observational data is available. We address two limitations of existing approaches for formal synthesis of controllers for networks of uncertain systems satisfying complex temporal specifications. Firstly, whilst existing approaches rely on the stochasticity to be Gaussian, the heterogeneous nature of composed systems typically yields a more complex stochastic behavior. Secondly, exact models of the systems involved are generally not available or difficult to acquire. To address these challenges, we show how abstraction-based control synthesis for uncertain systems based on sub-probability couplings can be extended to networked systems. We design controllers based on parameter uncertainty sets identified from observational data and approximate possibly arbitrary noise distributions using Gaussian mixture models whilst quantifying the incurred stochastic coupling. Finally, we demonstrate the effectiveness of our approach on
a nonlinear package delivery case study with a complex specification, and a platoon of cars.
\end{abstract}


\section{Introduction}
Cyber-physical systems (CPS) have become ubiquitous in almost all areas of modern life. Their adaptation in safety-critical areas, however, has lead to serious failures originating from the embedded controllers \cite{Axelrod2013CPSRisks}.
Designing CPSs that will not exhibit undesired or unsafe behavior 
when operating in an uncertain environment proves to be challenging.
%
Suppose we want to design a controller for an autonomous car. When driving in traffic, the car will perform various maneuvers among other vehicles and in possibly unseen scenarios. Whilst some knowledge of the dynamics of the car is usually available, exact models of the surrounding vehicles to which it is dynamically \emph{connected} are generally not available. 
Moreover, the \emph{heterogeneous} mixture of systems involved typically yields \emph{complex stochastic behavior}. 
How do we verify safe behavior if we are \emph{uncertain} about the environment and system and have to base our decisions on observations?
This is just one example of a typical problem arising in many domains of application.

Control synthesis for networks of stochastic systems to satisfy requirements expressed as temporal logic specifications is a challenging task. To obtain controllers with formal guarantees, a promising approach is to construct abstractions of the system and establish formal relations between the abstraction and the original system \cite{belta2017formal,lavaei2021automated,tabuada09}. Current abstraction-based approaches for networks of stochastic systems are limited in two main directions.
%
Firstly,
exact models of the systems in the network are usually not available or expensive to acquire.
Existing work, however, is mainly focused on systems with known mathematical models \cite{saoud2018composition,mallik2018compositional,lavaei2021compositional}.
%
Secondly,
the known stochastic behavior is assumed to be either bounded \cite{majumdar2023flexible} or of Gaussian nature \cite{haesaert2020robust}. 
Real-world examples of networked systems
exhibit a more complex stochastic behavior \cite{Blackmore2010ComplexDistr,Sun2006GMMTraffic}, e.g., due to a conglomerate of heterogeneous components.
This is a central feature disregarded in prior works on compositionality.

There is a limited body of work addressing systems with both stochastic and epistemic uncertainties.
%
Badings \textit{et~al.}~\cite{Badings2023NonGaussian} have studied monolithic linear systems with unknown additive noise 
for reach-avoid specifications. This work is extended in \cite{Badings2022epistemicUncert} to fully unknown linear systems.
In contrast, our approach can handle more complex specifications and nonlinear dynamics.
Compositional results for networks of unknown stochastic systems are provided in \cite{lavaei2022compositional} based on reinforcement learning, which rely on knowing the Lipschitz constants and are limited to finite-horizon specifications.
In the recent work \cite{Schoen2022ParamUncert}, we have shown that a relaxed version of stochastic simulation relations, called \emph{sub-simulation relations}, allows us to establish relations between uncertain stochastic systems and their abstractions.
However, the considered uncertainty is limited to the deterministic part of the dynamics whereas the stochastic behavior is assumed to be known a-priori. 
Moreover, it remains to be proved that these relations can similarly be applied to networked systems. 


In this work, we discuss the notion of sub-simulation relations for systems in a network. We show that such relations can be composed together to form a sub-simulation relation between networked systems.
We consider specifications that are conjunctions of local specifications defined on systems in the network.
To capture the expressive stochastic behavior of realistic networked systems, we construct surrogate models of systems with arbitrary additive noise distributions by approximating the noise using finite \emph{Gaussian mixture models} (GMM).
We show that the incurred error can be bounded even when the true noise distribution is unknown. 
This allows us to design controllers that are robust for networked systems subject to both stochastic and epistemic uncertainties.

\medskip

The paper is organized as follows. 
In Sec.~\ref{sec:prelim}, we give the preliminaries, introduce the class of models and specifications, and formulate a two-stage problem statement. 
Secs.~\ref{sec:SSRCompo} and \ref{sec:SSRGMM} are dedicated to providing solutions to these problems. In Sec.~\ref{sec:SSRCompo}, we provide the definition of sub-simulation relations for networks of systems and establish compositional results.
In Sec.~\ref{sec:SSRGMM}, we quantify the closeness for two systems with GMM noise distribution.
Finally, we demonstrate the proposed approach on a nonlinear package delivery case study and a platoon of cars in Sec.~\ref{sec:case_study}.

\section{Preliminaries and Problem Statement}\label{sec:prelim}
The following notation is used.
The transpose of a matrix $M$ is indicated by $M\T$.
Borrowing from common notation, for a column vector $x=[x^1;\ldots;x^n]\in\mathbb{R}^n$ we denote by $x^{-i}$ the vector deprecated by the $i^\text{th}$ element, namely $x^{-i}:=[ x^{1};\ldots;x^{i-1};x^{i+1};\ldots;x^{n}]$.
Similarly, we define the deprecated product of sets $\{A^i\}_{i=1}^n$ as
${A}^{-i}:=\prod_{j\neq i}A^j$. 

A measurable space is a pair $(\X,\mathcal{ F})$ with sample space $\X$ and $\sigma$-algebra $\mathcal{F}$ defined over $\X$,
which is equipped with a topology.
In this work, we restrict our attention to Polish sample spaces~\cite{bogachev2007measure}.
As a specific instance of $\mathcal F$, consider
Borel measurable spaces, i.e., $(\X,\mathcal{B}(\X))$, where $\mathcal{B}(\X)$ is the Borel $\sigma$-algebra on $\X$, that is the smallest $\sigma$-algebra containing open subsets of $\X$.
A positive \emph{measure} $\meas$ on $(\X,\mathcal{B}(\X))$ is a non-negative map
$\nu:\mathcal{B}(\X)\rightarrow \mathbb R_{\ge 0}$ such that for all countable collections $\{A_i\}_{i=1}^\infty$ of pairwise disjoint sets in $\mathcal{B}(\X)$
it holds that
$\nu\left({\bigcup_i A_i }\right)=\sum _i \nu({A_i})$.
A positive measure $\nu$ is called a \emph{probability measure} if $\nu(\X)=1$, and is called a \emph{sub-probability measure} if $\nu(\X)\leq 1$.

A probability measure $\po$ together with the measurable space $(\X,\mathcal{B}(\X))$ defines a \emph{probability space} denoted by $(\X,\mathcal{B}(\X),\po)$ and has realizations  $x\sim \po$.
We denote the set of all probability measures for a given measurable space $(\X,\mathcal{B}(\X))$ as $\mathcal P (\X)$.
For two measurable spaces $(\X,\mathcal{B}(\X))$ and $(\Y,\mathcal{B}(\Y))$, a \emph{kernel} is a mapping $\pok: \X \times \mathcal B(\Y)\rightarrow \mathbb R_{\geq 0}$
such that $\pk{x,\cdotx}:\mathcal B(\Y)\rightarrow\mathbb{R}_{\geq 0}$ is a measure for all $x\in\X$, and $\pk{\cdotx, B}: \X\rightarrow \mathbb R_{\geq 0}$ is measurable for all  $B\in\mathcal B(\Y)$.
A kernel associates to each point $x\in\X$ a measure denoted by $\pk{\cdotx|x}$.
We refer to $\pok$ as a \emph{(sub-)probability kernel} if in addition $\pk{\cdotx|x}:\mathcal B(\Y)\rightarrow [0,1]$ is a (sub-)probability measure.
The \emph{Dirac delta} measure $\delta_a:\mathcal{B}(\X)\rightarrow [0,1]$ concentrated at a point $a\in\X$ is defined as $\delta_a(A)=1$ if $a\in A$ and $\delta_a(A)=0$ otherwise, for any measurable $A\in\borel{\X}$.
The multivariate normal stochastic kernel with mean $\mu$ and covariance matrix $\Sigma$ is
denoted as $\N(dx|\mu, \Sigma)$.

For given sets $A$ and $B$, a relation $\rel\subset A\times B$ is a subset of the Cartesian product $A\times B$. The relation $\rel$ relates $x\in A$ with $y\in B$ if $(x,y)\in\rel$, written equivalently as $x\rel y$.
For a given set $\Y$, a metric or distance function $\dist$ is a function $\dist: \Y\times \Y\rightarrow \mathbb R_{\ge 0}$
satisfying the following conditions for all $y_1,y_2,y_3\in\Y$:
$\dist(y_1,y_2)=0$ iff $y_1=y_2$;
$\dist(y_1,y_2)=\mathbf d_{\Y}(y_2,y_1)$;  and
$\dist(y_1,y_3)\leq \dist(y_1,y_2) +\dist(y_2,y_3)$.

\subsection{Networks of uncertain stochastic systems}\label{sec:network}
In this work, we consider networked discrete-time uncertain nonlinear systems with two sources of uncertainty: (1) additive stochastic noise and (2) model parametric uncertainty.
Systems of this class can be represented by a model $\M(\theta)$ parametrized with $\theta=[\theta^1;\ldots;\theta^N]$ and
partitioned into $N$ subsystems $\M^i(\theta^i)$, $i\in\{1,\ldots,N\}$, as
\begin{equation}
	\label{eq:subsystem}
	\M^i(\theta^i): \left\{ \begin{array}{ll}
		x^i_{t+1}&= f^i(x^i_t,u^i_t;\theta^i) + w^i_t,   
		\\
		y^i_t &= h^i(x^i_t),
	\end{array} \right.
\end{equation}
where the state, input, and observation of the $i^{\text{th}}$ subsystem $\M^i(\theta^i)$ at the $t^{\text{th}}$ time-step are denoted by $x^i_t\in\X^i$, $u^i_t\in\U^i$, and $y^i_t\in\Y^i$, respectively. The state evolution and observation mapping are captured by the functions $f^i$ and $h^i$, respectively. The additive noise $w^i_t\in\mathbb{W}$ is an i.i.d. sequence with distribution $w^i_t\sim p^i_w(\cdotx|\theta^i)$.
Note that both the state evolution $f^i$ and the additive noise distribution $p^i_w$ are conditional on the uncertain parametrization $\theta^i$
for which we will assume an uncertainty set $\Theta$ such that $\theta\in\Theta$. This set can be constructed from observed input-output data with respect to a given confidence using system identification techniques as it is done in \cite{Schoen2023BayesSSR}.
%
We assume that the input and observation of the $i^{\text{th}}$ subsystem can be partitioned as $u_t^{i} = [u_t^{i1};\ldots;u_t^{iN}]$ and $y_t^{i} = [y_t^{i1};\ldots;y_t^{iN}]$, respectively.
The subsystems are dynamically linked via their \emph{internal} inputs and outputs $u_t^{-i}\in\U^{-i}$ and $y_t^{-i}\in\Y^{-i}$, respectively, as follows: $u_t^{ij} = y_t^{ji}, \forall i\neq j$.
We will refer to $u_t^{ii}\in\U^{ii}$ and $y_t^{ii}\in\Y^{ii}$ as \emph{external}.
For the network $\M(\theta)$, we recover the state, input, and observation as the concatenations $x_t=[x^1_t;\ldots;x^N_t]$, $u_t=[u^1_t;\ldots;u^N_t]$, $y_t=[y^1_t;\ldots;y^N_t]$.
%


\subsection{Gaussian mixture models}
We consider the noise distribution $p^i_w(w^i_t|\theta^i)$ to be a \emph{Gaussian mixture model} (GMM).
As a subclass of finite mixture models, GMMs are a widely used modeling framework for approximating probability distributions.
Apart from being particularly useful for capturing multiple sources of randomness, any continuous distribution can be approximated with arbitrary precision using GMMs \cite{McLachlan2000Mix}.

\begin{figure}
	\centering
	\includegraphics[width=.75\columnwidth]{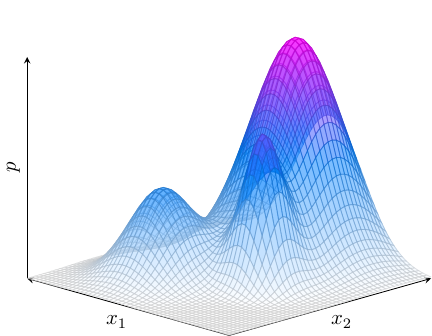}
	\caption{Example of a 2D Gaussian mixture with three components: $(\pi_1=0.2,\mu_1=[1;-1],\Sigma_1=0.5I)$, $(\pi_2=0.3,\mu_2=[-1;0],\Sigma_2=0.2I)$, and $(\pi_3=0.5,\mu_3=[0;2],\Sigma_3=I)$.}
	\label{fig:GMM}
\end{figure}

\begin{definition}[Gaussian Mixture Model (GMM)]
	A GMM is a probability measure $\GMM_K:\mathcal{B}(\X)\rightarrow[0,1]$ which is a weighted sum of finitely many ($K$) normal densities or \emph{component densities} $\N(x|\mu_k,\Sigma_k)$, $k\in\{1,\ldots,K\}$, i.e.,
	\begin{equation*}
		\GMM_K(dx|
		\pi,\mu,\Sigma
		) := \sum_{k=1}^{K} \pi_k \N(dx|\mu_k,\Sigma_k),
	\end{equation*}
	with \emph{mixing weights} $\pi:=(\pi_1,\ldots,\pi_K)$, $0\leq\pi_k\leq 1$, $\sum_{k=1}^{K}\pi_k=1$,
	mean values $\mu:=(\mu_1,\ldots,\mu_K)$, and covariance matrices $\Sigma:=(\Sigma_1,\ldots,\Sigma_K)$.
\end{definition}
A GMM is called \emph{homoscedastic} if all $K$ components share the same covariance matrix and \emph{heteroscedastic} otherwise.
Fig.~\ref{fig:GMM} depicts an example of a heteroscedastic GMM.
We omit the index indicating the number of components, i.e., $\GMM(\cdotx|\pi,\mu,\Sigma)$, when the number of components is uncertain.

For the networked system in \eqref{eq:subsystem}, the noise distributions can hence be written as $w^i_t\sim \GMM(\cdotx|\pi^i,\mu^i,\Sigma^i)$, where the parameters $(\pi^i,\mu^i,\Sigma^i)$ are contained in the unknown parametrization $\theta^i$. 

\subsection{Local control policies}
For each subsystem $\M^i$ in \eqref{eq:subsystem} initialized with an initial state $x^i_0$ at $t=0$ and an input sequence $u^i_0,u^i_1,u^i_2,\ldots$, consecutive states $x^i_{t+1}\in\X^i$ are obtained as realizations $x^i_{t+1}\sim f^i(x^i_t,u^i_t;\theta^i) + \GMM(w^i_t | \theta^i)$.
The \emph{execution history} $(x^i_0, u^i_0, x^i_1,\ldots, u^i_{N-1}, x^i_N)$ grows with the number of observations $N$ and takes values in the \emph{history space} $\Hist^i_N := (\X^i \times \A^i )^{N} \times \X^i$.
A local control policy or controller for $\M^i(\theta^i)$ is a sequence of policies mapping the current execution history to an external control input.
\begin{definition}[Local control policy]
	\label{def:markovpolicy}
	A local control policy $\boldsymbol{\pol}^i$ is a sequence $\boldsymbol{\pol}^i=(\pol^i_0,\pol^i_1,\pol^i_2,\ldots)$ of universally measurable maps $\pol^i_t:\Hist^i_t\rightarrow \mathcal P(\A^{ii},\mathcal B(\A^{ii}))$, $t\in\mathbb N:=\{0,1,2,\ldots\}$, from the execution history to a set of distributions on the external input space.
\end{definition}
As special types of control policies, we differentiate Markov policies and finite memory policies.
A \emph{Markov policy} $\boldsymbol{\pol}^i$ is a sequence $\boldsymbol{\pol}^i=(\pol^i_0,\pol^i_1,\pol^i_2,\ldots)$ of universally measurable maps $\pol^i_t:\X^i\rightarrow \mathcal P(\A^{ii},\mathcal B(\A^{ii}))$, $t\in\mathbb N$, from the state space $\X^i$ to a set of distributions on the external input space.
We say that a Markov policy is \emph{stationary}, if $\boldsymbol{\pol}^i=(\pol^i,\pol^i,\pol^i,\ldots)$ for some $\pol^i$.
\emph{Finite memory policies} first map the finite state execution of the system to a finite set (memory). The input is then chosen similar to the Markov policy as a function of the system state and the memory state. This class of policies is needed for satisfying temporal specifications on the system executions.
In the following, a local controller for each subsystem in \eqref{eq:subsystem} is denoted by $\Ca^i$ and the controlled subsystem by $\Ca^i\times \M^i$.

\subsection{Temporal logic specifications}
Consider a set of atomic propositions $AP := \{ p_1, \ldots, p_L \}$ that defines an \emph{alphabet} $\alphabeth := 2^{AP}$, where any \emph{letter} $\letter\in\alphabeth$ is composed of a set of atomic propositions.  An infinite string of letters forms a \emph{word} $\word=\letter_0\letter_1\letter_2\ldots\in\alphabeth^{\mathbb{N}}$.
We denote the suffix of $\word$ by $\word_j = \letter_j\letter_{j+1}\letter_{j+2}\ldots$ for any $j\in\mathbb N$.
Specifications imposed on the behavior of the system are defined as formulas composed of atomic propositions and operators. We consider the co-safe subset of linear-time temporal logic properties \cite{kupferman2001model} abbreviated as scLTL.
This subset of interest consists of temporal logic formulas constructed according to the following syntax
\begin{equation*}
	\psi ::=  p \ |\ \notltl p \ |\ \psi_1 \vee\psi_2  \ |\ \psi_1 \andltl \psi_2 \ |\ \psi_1 \Until \psi_2 \ |\ \Next \psi,
\end{equation*}   where $p\in \AP$ is an atomic proposition.
The \emph{semantics} of scLTL are defined recursively over $\word_j$ as
$\word_j \satisfies p$ iff $p \in \letter_j$;
$\word_j \satisfies \psi_1 \andltl  \psi_2  $ iff $ ( \word_j \satisfies \psi_1 ) \andltl ( \word_j \satisfies \psi_2 ) $;
$\word_j \satisfies \psi_1 \orltl  \psi_2  $ iff $ ( \word_j \satisfies \psi_1 ) \orltl ( \word_j \satisfies \psi_2 ) $;
$\word_j \satisfies  \psi_1 \Until \psi_2 $ iff $\exists m \geq j \text{ subject to } (\word_m \satisfies \psi_2 ) $ and $\word_t \satisfies \psi_1, \forall t \in \{j, \ldots m-1\}$; and
$\word_j \satisfies \Next \psi$ iff $\word_{j+1} \satisfies \psi$.
The eventually operator  $\Event \psi$ is used in the sequel as a shorthand for $\True\Until  \psi $.
We say that $\word\satisfies\psi$ iff $\word_0\satisfies\psi$.

Consider a labeling function $\mathcal{L}^i: \Y^i\rightarrow \Sigma^i$ of $\M^i$ that assigns a letter to each output. Using this labeling map, we can define temporal logic specifications over the output of the system.
Each output trace of the system $\mathbf{y}^i \!=\! y^i_0,y^i_1,y^i_2,\ldots$ can be translated to a word as $\word^i \!=\! \mathcal{L}^i(\mathbf{y}^i)$.
We say that a system satisfies the specification $\psi^i$ with the probability of at least $p_{\psi^i}$ if
$\mathbb P(\word^i\satisfies \psi^i) \ge p_{\psi^i}.$
When the labeling function $\mathcal{L}^i$ is known from the context, we write $\mathbb P(\Ca^i\times\M^i\satisfies \psi^i)$ to emphasize that the output traces of the controlled system $\Ca^i\times\M^i$ are used for checking the satisfaction.


\subsection{Problem statement}

In this work, we restrict ourselves to specifications that are decomposable as follows.
\begin{assumption}\label{asm:decompSpec}
	Let the system be decomposable into $N$ subsystems $\M^1,\ldots,\M^N$ and the global specification $\psi$ into $N$ local specifications $\psi^1,\ldots,\psi^N$ such that
	$ \psi = \bigwedge_{i=1}^{N}\psi^i$.
\end{assumption}
With this, we address networks of uncertain systems by solving the following two problems.
\begin{prob}
\label{prob:compo}
	Consider a networked uncertain system $\M(\theta)$ in \eqref{eq:subsystem} and specification $\psi$ satisfying Asm.~\ref{asm:decompSpec}.
	Let thresholds $p_{\psi^i}\in(0,1)$
	 for all $i\in\{1,\ldots,N\}$ be given.
	Design a global controller $\Ca$ with lower bound $p_\psi$ s.t.
	 \[ \P\left(\Ca\times \M(\theta) \satisfies \psi \right)\geq p_\psi, \quad \forall\theta\in\Theta = \prod_{i}\Theta^i,\]
from local controllers $\Ca^i$ for $\M^i(\theta^i)$ that satisfy \[\P\left(\Ca^i\times \M^i(\theta^i) \satisfies \psi^i\given \word^j\satisfies \psi^j,j\neq i\right)\geq p_{\psi^i}, \forall \theta^i\in\Theta^i.\]
\end{prob}
Then, for the local control synthesis we have the following.
%

\begin{prob}\label{prob:gmm}
	\setlength{\belowdisplayskip}{0pt}
	Given specification $\psi^i$ and threshold $p_{\psi^i}\in(0,1)$, design a local controller $\Ca^i$ for the model $\M^i(\theta^i)$ in \eqref{eq:subsystem} such that $\Ca^i$ is independent of $\theta^i$ and
	\[ \P\left(\Ca^i\times \M^i(\theta^i) \satisfies \psi^i \right)\geq p_{\psi^i}, \quad \forall\theta^i\in\Theta^i. \]
\end{prob}

As in \cite{Schoen2022ParamUncert}, we solve Prob.~\ref{prob:gmm} by 
computing a robust controller for a nominal system $\Mh$ in the set 
of feasible models \mbox{$\{\M(\theta)|\theta\in\Theta\}$}
and constructing a sub-simulation relation between the nominal controller and the set of models.
In particular, we design the coupling s.t. neither the interface function nor state refinement are dependent on the uncertain parametrization.
Note that here, in contrast to \cite{Schoen2022ParamUncert}, both the deterministic and stochastic part of the system dynamics are considered uncertain and the noise distribution is arbitrary.
In order to solve Prob.~\ref{prob:compo}, we synthesize local controllers by bounding the internal inputs and including the satisfaction of these bounds in the local specifications.
We derive a lower bound on the satisfaction probability of the networked system $\M(\theta)$ by showing that the local sub-simulation relations induce a global sub-simulation relation analogous to \cite{lavaei2021compositional}.
To solve Prob.~\ref{prob:gmm}, we provide results for establishing sub-simulation relations for systems with noise distributions in the form of GMMs.


\section{Sub-Simulation Relations for Networks of Uncertain Systems}
\label{sec:SSRCompo}

\begin{figure}[h]
	\centering
	\includegraphics[width=\columnwidth]{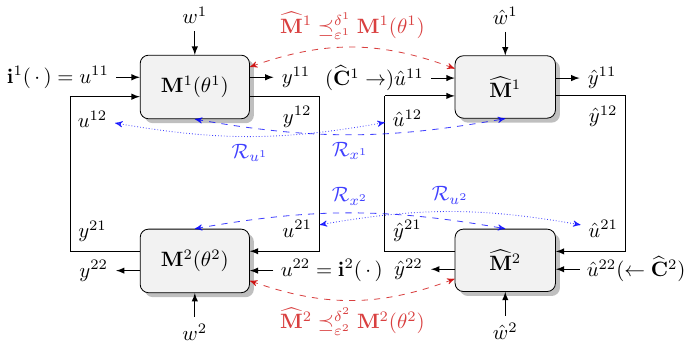}
	\caption{The network with parametric uncertainty (left) and its abstraction (right) are related by establishing simulation relations for each subsystem. These simulation relations quantify the similarity of the systems based on relations on the states and internal inputs.}
	\label{fig:SetupOverview}
\end{figure}

As in \cite{lavaei2021compositional}, our approach to Prob.~\ref{prob:compo} relies on constructing abstractions $\Mh^i$ of each individual subsystem $\M^i$ and designing local abstraction-based controllers $\Cah^i$. Based on local simulation relations between $\M^i$ and $\Mh^i$, the controllers $\Cah^i$ can be refined to controllers $\Ca^i$ for the original subsystems $\M^i$.
Fig.~\ref{fig:SetupOverview} illustrates the overall setup.
%

In this section, we show how the previous definition of sub-simulation relations extends to networks of uncertain systems and derive the corresponding compositional results.

\subsection{gMDPs and interconnections}
We represent the networked system in \eqref{eq:subsystem} as an interconnection of general Markov decision processes, defined next.
\begin{definition}[General Markov decision process (gMDP)]
	A gMDP is a tuple $\M\!=\!(\X,x_0,\A, \Tr, h,  \Y)$,
	with a state space $\X$ containing states $x\in\X$;
	an initial state $x_0\in\X$;
	an input space $\U$ with inputs $u\in\U$;
	a probability kernel $\Tr:\X\times\A\times\mathcal B(\X)\rightarrow[0,1]$;
	and an output space $\Y$ with a measurable output map $h:\X\rightarrow\Y$. The output space $\Y$ is decorated with a metric $\dist$.
\end{definition}
The \emph{transition} kernel $\Tr$ assigns to each state-input pair $(x,u)\in\X\times\U$ a probability measure $\Tr(\cdotx| x,u)$ on $(\X,\mathcal B(\X))$.
Hence, for a given system \eqref{eq:subsystem}, the probability of transitioning from a state $x$ with input $u$ to a state $\xp\in S$ is given by $\P(\xp\in S|x,u) = \int_{S} \Tr(d\xp|x,u;\theta)$, where the transition kernel can be written via the Dirac delta function as
\begin{equation*}
	\Tr(d\xp|x,u;\theta)= \int_{w}\delta_{f(x,u;\theta)+w}(d\xp) \; \GMM(dw | \theta).
\end{equation*}
The network of subsystems in \eqref{eq:subsystem} can hence be written as an interconnection of gMDPs.
We now provide the formal definition of an interconnection of $N$ subsystems described as gMDPs (cf. Fig.~\ref{fig:SetupOverview}).
\begin{definition}[Interconnection of subsystems]\label{def:interconnection}
Consider $N\in\mathbb{N}$ gMDPs $\M^i\!=\!(\X^i,x^i_0,\U^i, \Tr^i, h^i, \Y^i)$, $i\in\{1,\ldots,N\}$, with an input-output configuration that can be partitioned as outlined in Sec.~\ref{sec:network}.
Then, the \emph{interconnection} of the subsystems is itself a gMDP $\M\!=\!(\X,x_0,\U, \Tr, h, \Y)$, also denoted as $\Interc{\M^1,\ldots,\M^N}$, with $\X=\prod_{i}\X^i$, $\U=\prod_{i}\U^i$, $\Y=\prod_{i}\U^i$, and $h=[h^{11};\ldots;h^{NN}]$, and interconnection constraints
\begin{equation}
		\forall i\neq j:\quad u_t^{ij} = y_t^{ji}, \quad \Y^{ji}\subseteq\U^{ij}.
	\label{eq:interconnection_constraints}
\end{equation}
The kernel and initial state are given by $\Tr=\prod_{i}\Tr^{i}$ and $x_0=[x_0^{1};\ldots;x_0^{N}]$, respectively.
\end{definition}
Next, we provide the definition of sub-simulation relations for networked systems.

\subsection{Relation on subsystem level}
%
For each subsystem $\M^i$ in \eqref{eq:subsystem} we construct an abstract nominal model 
\begin{equation}
	\begin{split}
	\!\!\!\Mh^i: \left\{ \begin{array}{ll}
		\hat x^i_{t+1} \!\!\!\!\!&=  f^i(\hat x^i_t, \hat u^i_t;\hat\theta^i)+\hat w^i_t, \quad \hat w^i_t\sim \GMM(\cdotx | \hat\theta^i),\\
		\hat y^i_t &=  \hat h^i(x^i_t),
	\end{array} \right.\end{split}\label{eq:sysNom}
\end{equation}
with nominal parameters $\hat\theta^i$. 
Note that $\Mh^i:=\M^i(\hat\theta^i)$
 if $\hat h^i(\cdotx)= h^i(\cdotx)$. 
%
%
%
Based on the sub-probability coupling in \cite[Def.~5]{Schoen2022ParamUncert}, we establish sub-simulation relations for each subsystem pair $(\M^i(\theta^i),\Mh^i)$ that are independent of the concrete choice of the uncertain parametrizations $\theta^i$. 
The following definition extends the original definition of sub-simulation relations in \cite[Def.~6]{Schoen2022ParamUncert} to systems with both internal and external inputs.

%
%

\begin{definition}[$(\varepsilon^i,\delta^i)$-sub-simulation relation (SSR)]
	\label{def:ssr_subsys}
	Consider two gMDPs
	$\M^i\!=\!(\X^i,x^i_0,\U^i, \Tr^i, h^i, \Y^i)$ and $\Mh^i\!=\!(\Xh^i,\hat x^i_0,\Uh^i, \Trh^i, \hat h^i,  \Y^i)$, measurable relations $\Rx{i}\subset \Xh^i\times\X^i$ and $\Ru{-i}\subset {\Uh}^{-i}\times{\U}^{-i}$, and an interface function $\InF^i :\Xh^i\times\X^i\times\Uh^i\rightarrow\U^{ii}$.
	If there exists a sub-probability kernel $\Wt^i(\cdotx|\hat x^i,x^i,\hat u^i,u^{-i})$
	such that
	\begin{itemize}\itemsep=0pt
		\item[(a)] $(\hat x^i_0, x^i_0)\in \Rx{i}$;
		\item[(b)] \mbox{$\forall(\hat x^i,x^i)\in\Rx{i}$}, \mbox{$\forall\hat u^{i}\in\Uh^{i}$}
		 we have $u^{ii}=\InF^i(\hat x^i,x^i,\hat u^i)$ 
		such that \mbox{$\forall {u}^{-i}\in\Ru{-i}^{-1}\hat u^{-i}$}:
		$\Wt^i(\cdotx|\hat x^i,x^i,\hat u^{i}, {u}^{-i})$ is a sub-probability coupling  of $\hat \Tr^i(\cdotx|\hat x^i,\hat u^{i})$ and $\Tr^i(\cdotx|x^i,{u}^i)$ over $\Rx{i}$ with respect to $\delta^i$ (see \cite[Def.~5]{Schoen2022ParamUncert}); and
		\item[(c)] $\forall (\hat x^i,x^i)\in\Rx{i}: \mathbf \dist(\hat h^i(\hat x^i),h^i(x^i)) \leq \varepsilon^i$,
	\end{itemize}
	then $\Mh^i$ is in an $(\varepsilon^i,\delta^i)$-SSR with $\M^i$, denoted as \mbox{$\Mh^i\preceq^{\delta^i}_{\eps^i}\M^i$}.
\end{definition}


Intuitively, Def.~\ref{def:ssr_subsys} imposes three conditions on the composed system $\Mh^i\times\M^i$ evolving on the product space $\Xh^i\times\X^i$. They roughly correspond to its initial state, state transition, and output mapping (see Fig.~\ref{fig:SetupOverview}).
Condition (a) requires that the states of both systems start in $\Rx{i}$ upon initialization. Once in $\Rx{i}$, condition (b) certifies that for any external input  $\hat{u}^{ii}$ to $\Mh^i$ there exists a corresponding external input ${u}^{ii}$ to $\M^i$ such that the systems stay in $\Rx{i}$ with probability $(1-\delta^i)$ provided that the internal inputs of the two systems stay in relation $\Ru{-i}$. Note that the external input $\hat{u}^{ii}$ is mapped onto ${u}^{ii}$ by an interface function. Finally, according to condition (c), given that the two systems are in $\Rx{i}$, the corresponding outputs will be $\varepsilon^i$-close.

Def.~\ref{def:ssr_subsys} reduces to the original definition of $(\varepsilon,\delta)$-sub-simulation relations for monolithic systems \cite[Def.~6]{Schoen2022ParamUncert} if $i\in\{1\}$, or equivalently, $N=1$.


\medskip

In the next section, we prove that, as shown in \cite{lavaei2021compositional} for approximate simulation relations, local SSRs similarly induce a global SSR between the emerging interconnections.

\subsection{Relation on network level}
We now apply the introduced framework to networks subject to parametric uncertainty.
We establish an SSR \mbox{$\Mh^i\preceq^{\delta^i}_{\eps^i}\M^i$} between subsystems $(\M^i(\theta^i),\Mh^i)$ in Eqs.~\eqref{eq:subsystem} and \eqref{eq:sysNom}. We choose the interface function $u^{ii}=\InF^i(\hat x^i,x^i,\hat u^i):=\hat u^{ii}$ and the noise coupling 
\begin{align}\SwapAboveDisplaySkip
	& \hat w^i \equiv  \gamma^i(x^i, u^i,\theta^i;\hat\theta^i)+w^i, \,\,\text{ with an offset}\label{eq:noisecoup_subsys}\\
	& \gamma^i(x^i, u^i,\theta^i;\hat\theta^i) :=
	f^i(x^i,u^i;\theta^i)-f^i(x^i,u^i;\hat\theta^i),\label{eq:offset_subsys}
\end{align}
to get a state mapping which is not dependent on $\theta$:
\begin{equation}
	\xhp^i = \xp^i-f^i(x^i,u^i;\hat\theta^i )+ f^i(\hat x^i,\hat u^i;\hat\theta^i).\label{eq:statemap_subsys}
\end{equation}
Note that the noise coupling in \eqref{eq:noisecoup_subsys} implies that the distributions $\GMM(w|\theta)$, $\GMM(\hat w|\hat \theta)$ are of pairwise identical covariance, i.e., $\hat K=K$ and $\hat\Sigma_k=\Sigma_k, \,\forall k\in\{1,\ldots,K\}$. This does not, however, limit the expressiveness of the noise distributions that can be addressed and is more general than assuming homoscedasticity.
The following choice of relations enables us to find a convenient formulation of the coupling:
\begin{align}
	\Rx{i}&:=\big\lbrace(\hat x^i,x^i)\in\Xh^i\times\X^i\given\hat x^i=x^i\big\rbrace\label{eq:relation_subsys},\\
	\Ru{-i}&:=\big\lbrace(\hat u^{-i},u^{-i})\in\Uh^{-i}\times\U^{-i}\given\hat u^{-i}=u^{-i}\big\rbrace\label{eq:relationinput_subsys},\\
	\Rx{} &:= \big\lbrace(\hat x, x)\in\Xh\times\X\given\hat x=x\big\rbrace,\label{eq:relation_intercon}\\
	&\equiv\big\lbrace(\hat x, x)\in\Xh\times\X\given\forall i\in\{1,2\}:\; (\hat x^i, x^i)\in\Rx{i}\big\rbrace\nonumber.
\end{align}


Condition (a) of Def.~\ref{def:ssr_subsys} holds by setting the initial states $\hat x^i_0 = x^i_0$, $i\in\{1,\ldots,N\}$.
For condition (b), we define the sub-probability coupling $\Wt^i$ of $\Trh^i$ and $\Tr^i$ over $\Rx{i}$ as 
\begin{align}\SwapAboveDisplaySkip
		&\Wt^i(d\xhp^i\!\times \!d\xp^i|\theta^i) 
		= \!\!\int_{w^i} \!\!\delta_{f^i(\hat\theta^i)+ w^i+\offset^i}(\dxhp^i) \delta_{f^i(\theta^i)+ w^i}(\dxp^i)\hspace{10pt}
		\nonumber\\
		&\hspace{10pt}
		 \min\big\lbrace
		\GMM_{K^i}(dw^i|\theta^i), 
		\GMM_{K^i}(dw^i|\hat\pi^i,\hat\mu^i-\offset^i, \Sigma^i)
		\big\rbrace\!
	\label{eq:subprobcoup_subsys},
\end{align}
where we dropped the dependence on $(x^i, u^i)$ for brevity. 
Notice that the marginals of the coupling $\Wt^i$ represent a lower-bound on the transition kernels of $\M^i(\theta^i)$ and $\Mh^i$.

\begin{theorem}[\bfseries Induced compositional SSR]\label{thm:comp_induction_two}
	Let gMDPs
	$\{\M^i\}_{i=1}^N$ and $\{\Mh^i\}_{i=1}^N$
	as in \eqref{eq:subsystem} and \eqref{eq:sysNom} be given with
	common metric $\dist=\norm{\cdotx}$,
	SSRs $\Mh^i\preceq^{\delta^i}_{\eps^i}\M^i$ for $i\in\{1,\ldots,N\}$ with relations $\Rx{i}$ and $\Ru{-i}$ in \eqref{eq:relation_subsys}-\eqref{eq:relationinput_subsys}, sub-probability couplings $\Wt^i$ in \eqref{eq:subprobcoup_subsys}, and interface functions $u^{ii}=\hat u^{ii}$. Furthermore, let 
	$\M\!=\!
	\Interc{\M^1,\ldots,\M^N}$ and 
	$\Mh\!=\!
	\mathscr{I}(\Mh^1,\ldots,\Mh^N)$ be the corresponding interconnections.
	If $\forall i\in\{1,\ldots,N\}$ with $ j\neq i$ and $\, \hat u^{ij} \equiv h^{ji}(x^j)$ we have
	\begin{equation*}
			\forall (\hat x^i, x^i)\in\Rx{i}: \; (\hat u^{-i}, u^{-i})\in\Ru{-i},
	\end{equation*}
	then $\Mh$ is in an \emph{induced compositional} $(\varepsilon, \delta)$-SSR with $\M$, denoted  \mbox{$\Mh\preceq^{\delta}_\eps\M$} with relation $\Rx{}$ in \eqref{eq:relation_intercon}
	and $\varepsilon=\sum_i\varepsilon^i$, $\delta=1-\prod_{i}(1-\delta^i)$. Furthermore, we obtain the sub-probability coupling $\Wt=\prod_{i}\Wt^{i}$ and the interface function $\InF=\prod_{i}\InF^{i}$, applying the interconnection constraints \eqref{eq:interconnection_constraints}.
\end{theorem}
The proofs of statements have been deferred to the appendix.
Given the results of Thm.~\ref{thm:comp_induction_two}, we proceed to providing global guarantees for the networked system.

\subsection{Global guarantees}
	The probability of a local controlled subsystem $\Ca^i\times\M^i$ satisfying the specification $\psi^i$ is conditioned on 
	the other subsystems.
	We obtain the probability that the interconnected system $\Ca\times\M$ satisfies the global specification $\psi$ as the intersection of events $(\Ca^i\times\M^i\satisfies\psi^i)$.
	
We establish results for two different network structures. 
In a \emph{cascaded} network, the interconnection between subsystem is unidirectional and every subsystem is only influenced by its predecessors. In a \emph{cyclic} network, there is at least one cycle in the interconnection graph of the subsystems.
Let us index the predecessors of the $i^{\text{th}}$ subsystem  in the network via $\mathrm{Pre}(i)$.

\begin{theorem}[Global guarantees (cascaded)]\label{thm:globalGuaranteesCascaded}
	Consider a cascaded network of local controlled subsystems $\Ca^i\times\M^i$. The global probability of satisfaction w.r.t. $\psi$ is lower bounded by\\[-.5em]
\begin{equation}
\label{eq:globalGuaranteesCascaded} 
		\P\!\left(\Ca\!\times\!\M\!\satisfies\!\psi\right)\! \geq \!
		\prod_{i=1}^{N} \!\!\!\!\!\min_{\hspace{.9em}\mathbf{y}^j\in\mathbf{Y}^j}\!\!\!\!\! \P\Big(\Ca^i\times\M^i\!\satisfies\!\psi^i \,|\, \mathbf{y}^j, j\in\mathrm{Pre}(i)\Big),
	\end{equation}
	where $\mathbf{Y}^j:=\big\lbrace\mathbf{y}^j:\mathcal{L}^j(\mathbf{y}^j)\satisfies\psi^j\big\rbrace$.
\end{theorem}

For a general \emph{circular} network configuration, i.e., 
when there are feedback loops or cycles in the interconnection graph of the subsystems, we obtain a more conservative lower bound. 
For this, we bound the internal outputs of the subsystems, i.e., $\mathcal{C}^i\subset\Y^i$ for all $\smash{i\in\{1,\ldots,N\}}$, and assign an output label $p_{\mathcal{C}^i}$.
We define the event of a subsystem satisfying this bound as the safety specification $\mathcal A^i:=(\Ca^i\times\M^i\satisfies\square p_{\mathcal{C}^i})$.
We obtain the probability that the interconnected system $\Ca\times\M$ satisfies the global specification $\psi$ as the intersection of events $(\Ca^i\times\M^i\satisfies\psi^i\wedge\mathcal{A}^i)$.
\begin{theorem}[Global guarantees (cyclic)]\label{thm:globalGuaranteesCircular}
	Consider a cyclic network of locally controlled subsystems $\Ca^i\times\M^i$. The global probability of satisfaction w.r.t. $\psi$ is lower bounded by\\[-.5em]
	\begin{equation}
		\P\!\left(\Ca\!\times\!\M\!\satisfies\!\psi\right)\! \geq \!
		\prod_{i=1}^{N} \!\!\!\!\!\!\min_{\hspace{.9em}\mathbf{y}^j\in\mathbf{Y}^j}\!\!\!\!\! \P\Big(\!(\Ca^i\times\M^i\!\satisfies\!\psi^i)\andltl\mathcal{A}^i | \mathbf{y}^j\!\!, j\!\in\!\mathrm{Pre}(i)\!\Big)
		,\label{eq:globalGuaranteesCircular}
	\end{equation}
	where $\mathbf{Y}^j:=\big\lbrace\mathbf{y}^j:\mathcal{L}^j(\mathbf{y}^j)\satisfies\mathcal{A}^j\big\rbrace$.
\end{theorem}
\section{Sub-Simulation Relations for Gaussian Mixture Models}\label{sec:SSRGMM}
Now that we can compute global guarantees on the networked system based on local guarantees on the individual subsystems $\M^i$, we solve Prob.~\ref{prob:gmm} by showing how to establish an SSR (Def.~\ref{def:ssr_subsys}) for a system with GMM noise as in \eqref{eq:subsystem}.
Note that we consider the noise to be uncertain as well.
 In the following, we consider every subsystem individually and hence drop the exponent $i$.

\medskip

\begin{theorem}[GMM $\delta$]\label{thm:gmmDelta}
	The subsystems in \eqref{eq:subsystem} and \eqref{eq:sysNom} are in an SSR
	$\widehat{\M}\preceq^{\delta}_\eps\M(\theta)$ with interface function $\smash{\ac{k} = \ach{k}}$, relation \eqref{eq:relation_subsys}
	, and sub-probability coupling \eqref{eq:subprobcoup_subsys}
	.
	The state mapping \eqref{eq:statemap_subsys}
	 defines a valid control refinement with $\smash{\eps=0}$ and
	\begin{align}\SwapAboveDisplaySkip
			&\delta(x, u) = 
			\sup_{\theta\in\Theta}\left\lbrace 1-
			\!\sum_{k=1}^{K}\! \left[ \pi_k \cdf{\!-\!\left(\dfrac{1}{2}-\eta_k\right)\!\norm{\beta_k}}\right.\right.		\label{eq:deltafcn}\\
			&\hspace{90pt}+ \left.\left.\hat\pi_k \cdf{\!-\!\left(\dfrac{1}{2}+\eta_k\right)\!\norm{\beta_k}} \right]
			\right\rbrace, \nonumber
	\end{align}
	with coefficients
	\begin{align*}
		\eta_k(x,u,\theta) &:= \dfrac{1}{\beta_k(x,u,\theta)\T\beta_k(x,u,\theta)}\log\dfrac{\pi_k}{\hat\pi_k},\quad\text{and}\\
		\beta_k(x,u,\theta) &:= \mu_k-\hat\mu_k+\offset(x,u,\theta),
	\end{align*}
	where $\cdf{\cdotx}$ denotes the cumulative distribution function of a Gaussian distribution, i.e., $\cdf{\zeta}:= \int_{-\infty}^\zeta \frac{1}{\sqrt{2\pi}}\exp(-\beta^2/2)d\beta$, and offset $\offset$ as defined in \eqref{eq:offset_subsys}
	.
\end{theorem}

Thm.~\ref{thm:gmmDelta} extends the results from \cite{Schoen2022ParamUncert} to more general noise distributions. The results from \cite[Thm.~4]{Schoen2022ParamUncert} are recovered for $\mu_k = \hat{\mu}_k$ and $K=1$ (implying $\pi_k=\hat{\pi}_k=1$).

\section{Case Studies}
\label{sec:case_study}
In this section, we demonstrate the capabilities of the proposed extensions on a monolithic package delivery case study and a platoon of cars with two coupled subsystems.

\subsection{Package delivery}

\begin{figure}
	\centering
	\includegraphics[width=\columnwidth]{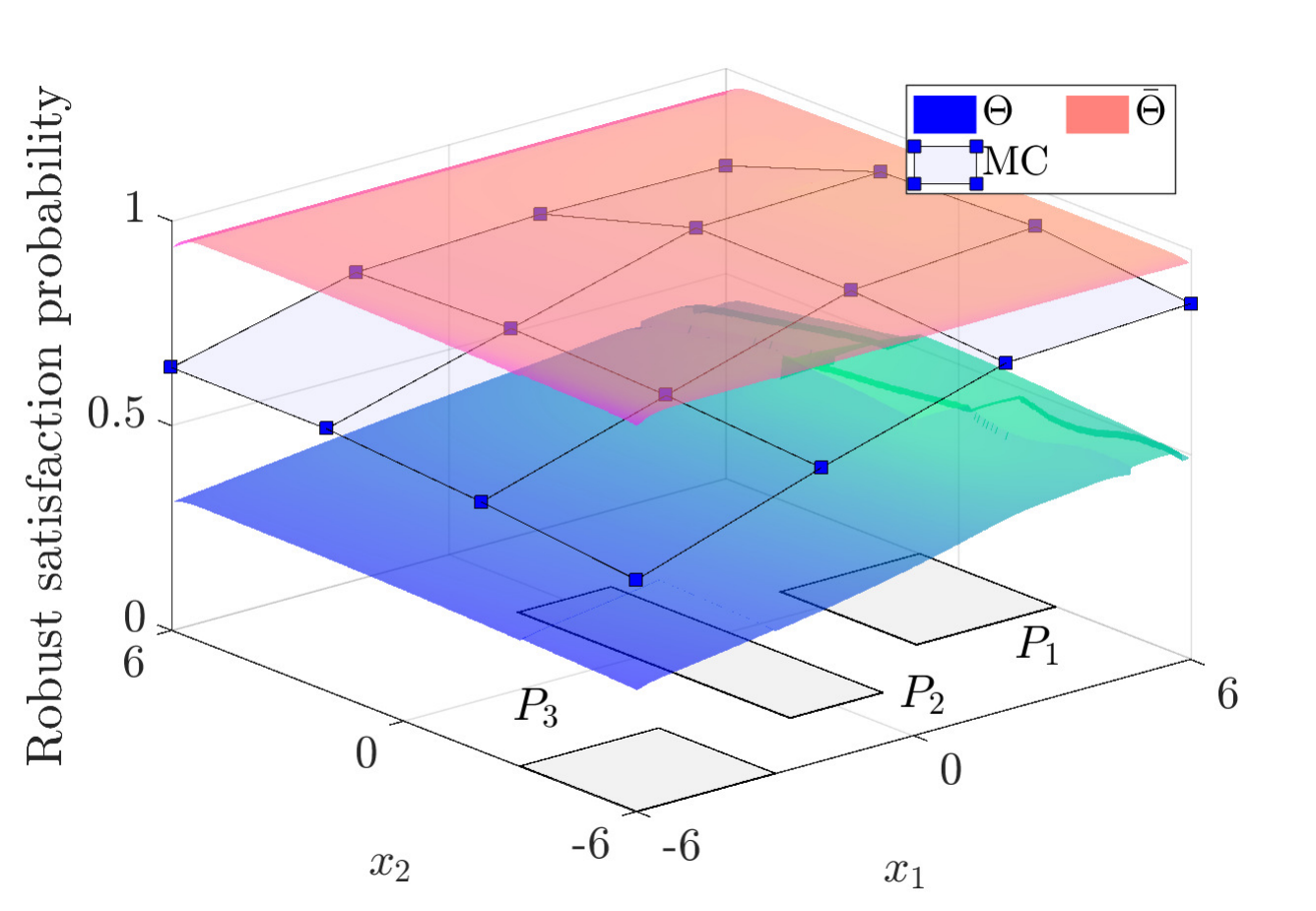}
	\caption{Lower bound on the satisfaction probability as a function of the initial state and the actual satisfaction probability estimated via Monte Carlo simulation for the package delivery case study for $\Theta$ in blue. The robust lower bound is boosted for a contracting uncertainty set ($\bar\Theta$ in red).}
	\label{fig:satProb_PkgDel}
\end{figure}

Consider a monolithic uncertain nonlinear system
\begin{equation*}
	\M(\theta): \left\{ \begin{array}{ll}
		\!\! x_{t+1}\!\!\!\!\!\!&= \begin{bmatrix}0.9x^1_t+0.6\sin(0.1x^2_t)+1.7\bar\theta u_t^1\\ 0.9x_t^2+1.7u_t^2\end{bmatrix}
		+ w_t, 
		\\
		\!\! y_t &= x_t,
	\end{array} \right. 
\end{equation*}
with state $x_t=[x_t^1;x_t^2]$ and input $u_t=[u_t^1;u_t^2]$,
describing an agent translating in a 2D space.
Note that here, the superscript `$1$', `$2$' refers to the elements of $x_t$ and $u_t$.
The goal is to compute a controller to navigate the agent for collecting a parcel in region $P_1$ and delivering it to target region $P_3$. If the agent visits the avoid region $P_2$ on its path, it loses the package and must collect a new one at $P_1$. This behavior is captured by the specification $\psi =\!\!\Event\!(P_1 \andltl (\notltl P_2 \Until P_3))$. 
	Note that this is a complex specification that cannot be expressed as a reach-avoid specification.
	The regions are given on the $xs$-plane in Fig.~\ref{fig:satProb_PkgDel}.
The homoscedastic noise distribution $w_t\sim\GMM_2(\cdotx|\pi,\mu,\Sigma)$ has the 
common covariance matrix
$\Sigma:=(\bar\Sigma,\bar\Sigma)$, $\bar\Sigma=[\sqrt{0.2},0;0,\sqrt{0.2}]$.
The uncertain parameterization is $\theta:=(\bar\theta,\pi,\mu)$ with $\bar\theta=1$.
%
We define the state space $\X = [-6,6]^2$, input space $\U = [-1,1]^2$, and output space $\Y = \X$.

The uncertainty set $\Theta$ has the elements
$\bar\theta\in[0.950,1.050]$,
$\pi=(0.8,0.2)$,
$\mu_1\in[-0.010, 0.010]\times[0.790, 0.810]$, and
$\mu_2\in[-0.810, -0.790]^2$.
We select a nominal model 
$\Mh=\M(\hat\theta)$ 
where $\hat\theta$ has the elements $\bar\theta=0.99$, $\pi=(0.8,0.2)$, $\mu_1=[0;0.8]$, and $\mu_2=[-0.8;-0.8]$,
and get $\eps_1=0$. 
An input-dependent $\delta_1$ is computed using Eq.~\eqref{eq:deltafcn}. 
We compute a second abstract model $\Mt$ by discretizing the space of $\Mh$
using the method outlined in \cite{huijgevoort2022piecewiseaffineabstraction}.
Then, we use the results therein to get ${\Mt}\preceq^{\delta_2}_{\eps_2}\Mh$ with $\eps_2=0.060$ and a state-dependent $\delta_2$. Thus, using the transitivity property in \cite[Thm.~2]{Schoen2022ParamUncert}, we have $\Mt\preceq^{\delta}_{\eps}\M(\theta)$ with $\delta = \delta_1+\delta_2$ and $\eps = \eps_1 + \eps_2$.
We compare the robust probability of satisfying the specification computed using this $\eps$ and $\delta$ based on \cite[Prop.~1]{Schoen2022ParamUncert} with the true satisfaction probability for a parametrization 
$\theta$ with the elements $\bar\theta=1$, $\pi=(0.8,0.2)$, $\mu_1=[0;0.8]$, and $\mu_2=[-0.8;-0.8]$,
estimated using Monte Carlo simulation for several representative initial states. We run $10^5$ simulations per initial state with a maximum length of 30 time steps. Fig.~\ref{fig:satProb_PkgDel} shows the robust satisfaction probability (in blue) as a function of the initial state of $\M(\theta)$ alongside the actual satisfaction probability (blue mesh) estimated via Monte Carlo simulation.
Moreover, Fig.~\ref{fig:satProb_PkgDel} shows (in red)
the robust satisfaction probability of a controller synthesized for a tighter uncertainty set
$\bar\Theta$ with the elements $\bar\theta\in[0.999,1.001]$, $\pi=(0.8,0.2)$, $\mu_1\in[-0.001, 0.001]\times[0.799, 0.801]$, and $\mu_2\in[-0.801, -0.799]^2$,
obtained using a bigger data set. 

\begin{figure*}[bth!]
	\centering
	\includegraphics[width=.66\columnwidth]{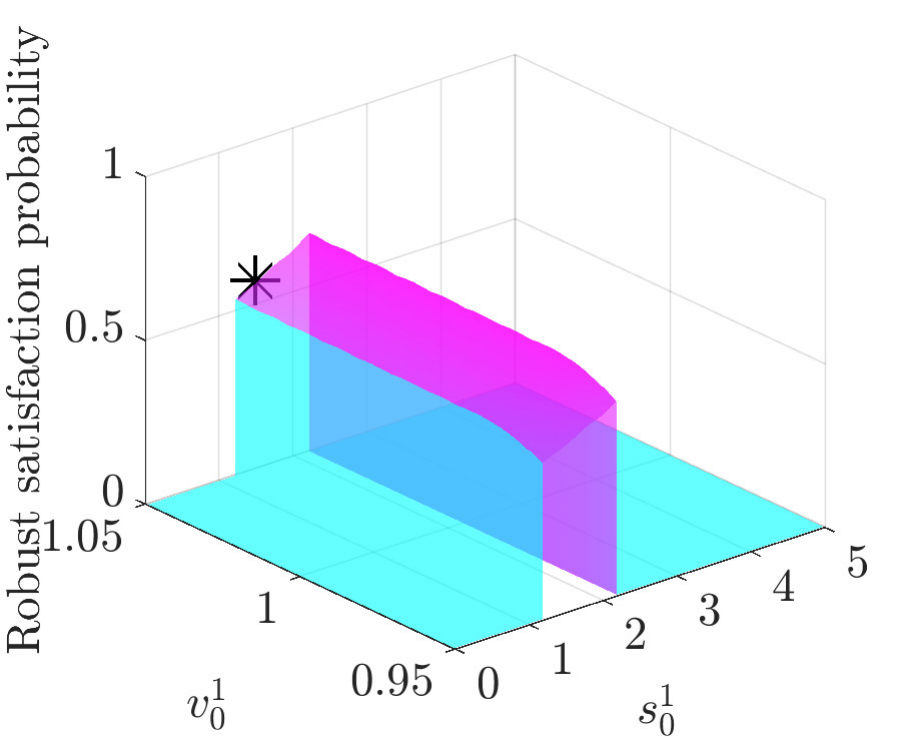}
	\includegraphics[width=.66\columnwidth]{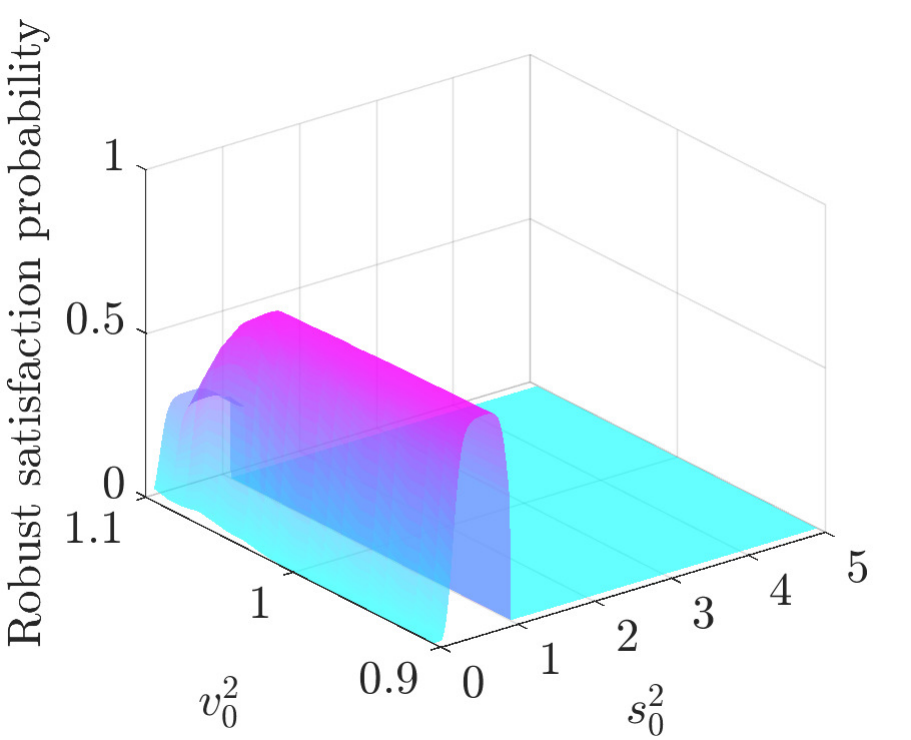}
	\includegraphics[width=.66\columnwidth]{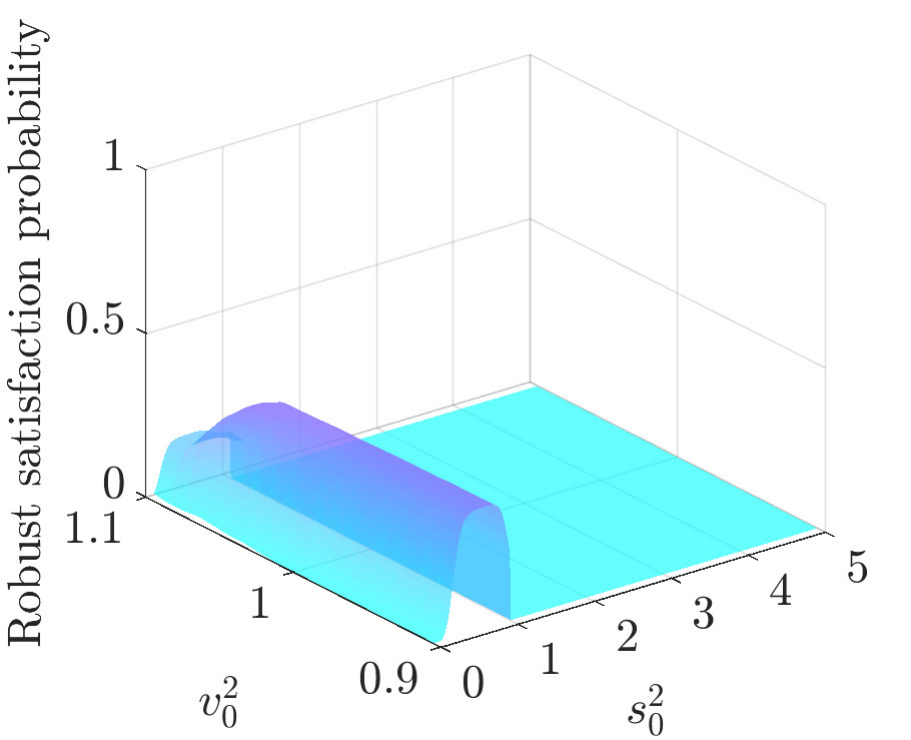}
	\caption%
	{Lower bound on the local satisfaction probability as a function of the initial state of the leading vehicle left and following vehicle (middle), as well as a lower bound on the global satisfaction probability as a function of the initial state of the following vehicle (right) with the leading car initialized at $[s^1_{0};v^1_{0}]=[1.5;1.05]$. 
	}
	\label{fig:satProb_CP}
\end{figure*}

\subsection{Car platoon}
Now, we look at a networked system consisting of two subsystems. In particular, we consider a platoon of two cars driving behind each other.
The dynamics of the \emph{leading vehicle} $\M^1(\theta^1)$ are given by
\begin{equation*}
	\begin{bmatrix}s^1_{t+1}\\v^1_{t+1}\end{bmatrix} = \begin{bmatrix}
		1 &  \tau \\
		0 & 0.9
	\end{bmatrix} \begin{bmatrix}s^1_{t}\\v^1_{t}\end{bmatrix} + \begin{bmatrix}0\\\frac{\tau}{m^1}\end{bmatrix}u^1_t + w^1_t,
\end{equation*}
with 
constant $\tau=0.5$,
output mapping $y^1_t=x^1_t$, 
and a homoscedastic noise distribution $w^1_t\sim\GMM_2(\cdotx|\pi^1,\mu^1,\Sigma^1)$ with common covariance matrix $\bar\Sigma^1:=[0.050,0;0,0.025]$, i.e., $\Sigma^1 := (\bar\Sigma^1,\bar\Sigma^1)$. 
The uncertain parameters are $\theta^1:=(m^1,\,\pi^1,\,\mu^1)$.
The state, input, and output spaces are $\X^1 = [0,5]\times[0.95,1.05]$, $\U^1 = [0.3,1]$, and $\Y^1 = \X^1$, respectively.
%
The dynamics of the \emph{following vehicle} $\M^2(\theta^2)$ are
\begin{equation*}
	\begin{bmatrix}s^2_{t+1}\\v^2_{t+1}\\d_{t+1}\end{bmatrix} = \begin{bmatrix}
		1 & 1.8\tau & 0\\
		0 & 0.9 & 0\\
		0 & -\tau & 1
	\end{bmatrix} \begin{bmatrix}s^2_{t}\\v^2_{t}\\d_t\end{bmatrix} + \begin{bmatrix}0\\\frac{\tau}{m^2}\\0\end{bmatrix}u^2_t
 + \begin{bmatrix}0\\0\\\tau\end{bmatrix}v^1_t 
 + w^2_t,
\end{equation*}
where the third state variable couples the two subsystems via $v^1_t$, capturing the distance between the vehicles.
Similarly as before, we have the
output mapping $y^2_t=x^2_t$,
and a noise distribution $w^2_t\sim\GMM_2(\cdotx|\pi^2,\mu^2,\Sigma^2)$
with common covariance matrix $\bar\Sigma^2:=[0.075,0,0;0,0.025,0;0,0,0.050]$, $\Sigma^2 := (\bar\Sigma^2,\bar\Sigma^2)$.
Note how the GMM allows to capture the sensor noise on the velocity measurements of $\M^1(\theta^1)$.
The uncertain parameters are $\theta^2:=(m^2,\,\pi^2,\,\mu^2)$.
Define the state space $\X^2= [0,5]\times[0.9,1.1]\times[0.2,2.2]$, input space $\U^2 = [0.5,1.5]$, and output space $\Y^2 = \X^2$.

The global specification can be decomposed as follows.
The goal for $\M^1$ is to start in initial region $P^1_{\text{init}}:=[1.2,2.2]\times[0.95,1.05]$ and reach target region $P^1_{\text{targ}}:=[4,5]\times[0.95,1.05]$ whilst remaining in $P^1_{\text{safe}}:=[1.2,5]\times[0.95,1.05]$, written as $\psi^1:=P^1_{\text{init}}\andltl \Next(P^1_{\text{safe}}\Until P^1_{\text{targ}})$.
Similarly, we define $\psi^2:=P^2_{\text{init}}\andltl \Next(P^2_{\text{safe}}\Until P^2_{\text{targ}})$ with $P^2_{\text{init}}:=[0,1]\times[0.9,1.1]\times[0.2,2.2]$, $P^2_{\text{targ}}:=[2.8,3.8]\times[0.9,1.1]\times[0.2,2.2]$, and $P^2_{\text{safe}}:=[1,3.8]\times[0.9,1.1]\times[0.2,2.2]$.
Note that $(\psi^1,\psi^2)$ bound the velocities via $v^1_t\in\mathcal{C}^1:=[0.95,1.05]$ and $v^2_t\in\mathcal{C}^2:=[0.9,1.1]$, $\forall t\geq 0$, respectively.

The uncertainty sets $\hat\Theta^i$ are given as $\hat\Theta^1$ with the elements $m^1\in[3.9,4.1]$, $\pi^1=(0.3, 0.7)$, $\mu^1\in([0.09, 0.11]\times[-0.01, 0.01],[-0.01, 0.01]\times[-0.01, 0.01])$,
and $\hat\Theta^2$ with the elements $m^2\in[3.8,4.0]$, $\pi^2=(0.3,0.7)$, $\mu^2\in([-0.010, 0.010]^2\times[0.009, 0.011], [-0.010, 0.010]^2\times[-0.011, -0.009])$.
We select nominal models $\Mh^i=\M^i(\hat\theta^i)$
based on $\hat\theta^1$ with the elements $m^1 = 4$, $\pi^1=(0.3, 0.7)$, $\mu^1 = ([0.1; 0], [0; 0])$, and $\hat\theta^2$ with the elements $m^2 = 3.9$, $\pi^2=(0.3, 0.7)$, $\mu^2 = ([0; 0; 0.01], [0; 0; -0.01])$,
and get $\eps^i_1=0$.
Input-dependent $\delta^i_1$ are computed using Eq.~\eqref{eq:deltafcn}. 
%
We construct a second batch of abstract models $\Mt^i$ by discretizing the space of $\Mh^i$.
Then, we use the results of \cite{haesaert2020robust} to get ${\Mt^i}\preceq^{\delta_2^i}_{\eps_2^i}\Mh^i$ with $(\eps^1_2=0.006, \delta^1_2=0.002)$ and $(\eps^2_2=0.030, \delta^2_2=0.007)$.
Note that we account for the error inflicted by the internal input by augmenting the discretization error in the invariance constraint \cite[Eq.~(20d)]{huijgevoort2020similarityquantification}.
Using the transitivity property in \cite[Thm.~2]{Schoen2022ParamUncert}, we have \mbox{$\Mt^i\preceq^{\delta^i}_{\eps^i}\M^i(\theta^i)$} with $\delta^i=\delta_1^i+\delta_2^i$ and $\eps^i=\eps_1^i+\eps_2^i$.
%
%
To synthesize local controllers, we bound $(v^1_{t}- v^2_{t})\in[-0.15,0.15]$ and take the worst case in each step.
The robust probability of each subsystem satisfying their respective specification is computed based on \cite[Prop.~1]{Schoen2022ParamUncert} and is depicted in Fig.~\ref{fig:satProb_CP} (left and middle) as functions of the initial state of $\M^1(\theta^1),\M^2(\theta^2)$.
The global satisfaction probability of the network is computed based on the individual probabilities using Eq.~\eqref{eq:globalGuaranteesCascaded} and is depicted in Fig.~\ref{fig:satProb_CP} (right) as a function of the initial state of $\M^2(\theta^2)$ for $\M^1(\theta^1)$ initialized at $[s^1_{0};v^1_{0}]=[1.5;1.05]$.

\section{Conclusion}
In this paper, we extended the definition of sub-simulation relations to establish quantitative relations between parametrized networks of systems and their abstractions. Moreover, we provided results for systems with complex stochastic behavior by approximating the uncertain noise distribution using Gaussian mixture models. We demonstrated the extensions on two intricate case studies.
In the future, we plan to address systems with noisy observations and partially observable systems. 
 
\bibliographystyle{abbrv}
\bibliography{references.bib}

\appendix

\subsection{Proof of Thm.~\ref{thm:comp_induction_two}}
\begin{proof}
	We show that $\Mh$ is in an SSR with $\M$ by proving that the conditions in Def.~\ref{def:ssr_subsys} hold, for the monolithic case of $i\in\{1\}$ (equivalent to \cite[Def.~6]{Schoen2022ParamUncert}).
	Condition (a) holds by setting the initial states $\hat x_{0} = x_{0}$.
	%

\noindent\emph{$\varepsilon$-deviation:}
The proof of condition (c) is a trivial extension to the one given for \cite[Thm.~4.3]{lavaei2021compositional}. \newline 
\noindent\emph{$\delta$-deviation:}
For condition (b), we complete the sub-probability coupling $\Wt^i$ in \eqref{eq:subprobcoup_subsys} to a probability kernel via
\begin{align}
		&\fWt^i (d\xh{}^i\times d\x{}^i) = \sWt^i(d\xh{}^i\times d\x{}^i) +  \tfrac{1}{1-\sWt^i(\Rx{i})} \label{eq:fullcoupling}
		\\&\hspace{35pt}
		\big(\Tr(d\x{}^i)-\sWt^i(\Xh^i \times d\x{}^i)\big)\big(\Trh^i(d\xh{}^i)-\sWt^i(d\xh{}^i \times \X^i)\big),\nonumber
\end{align}
omitting the argument $\theta^i$ for conciseness.
Reference \cite[Thm.~1]{Schoen2023BayesSSR} for more details.
We establish the following lemma.
\begin{lemma}\label{lem:maxcoup}
	The sub-probability coupling $\Wt^i$ in Eq.~\eqref{eq:subprobcoup_subsys} is maximal w.r.t. transition kernels $(\Trh^i,\Tr^i)$ in \eqref{eq:subsystem} and \eqref{eq:sysNom} and relation $\Rx{i}$ in \eqref{eq:relation_subsys}, i.e., for the corresponding probability kernel $\fWt^i$ \cite[Thm.~1]{Schoen2023BayesSSR} we have $\fWt^i(\Rx{i})=\Wt^i(\Rx{i})$.
\end{lemma}
\begin{proof}
	In the following, we drop the superscript $i$ for clarity of notation. 
	We integrate $\fWt$ in \eqref{eq:fullcoupling} over $\Rx{}$ and get
	\begin{align}\SwapAboveDisplaySkip
			\Wt(\Rx{}) &= \fWt(\Rx{}) - \tfrac{1}{1-\Wt(\Rx{})} \phi(\Rx{}),\quad\text{where}\label{eq:fullcoup_temp}\\
			\phi(\Rx{}) &:= \int_{\X} \big(\Tr(d\x{})-\Wt(\Xh \times d\x{})\big)\big(\Trh(d\x{})-\Wt(d\x{} \times \X)\big).\nonumber
	\end{align}
	Using the definitions of $\Wt,\Tr$, and $\Trh$ we expand and get 
	\begin{equation*}
		\phi(\Rx{}) \!=\! \int_{\X}
		\int_{\mathbb{W}}\!\delta_{f(\hat\theta)+w}(dx) p_w(dw)
		\int_{\mathbb{W}}\!\delta_{f(\hat\theta)+\hat w}(dx) p_{\hat w}(d\hat w),
	\end{equation*}
	with distributions 
	\begin{align*}\SwapAboveDisplaySkip
		p_w(dw) &:= \min\big\lbrace0,\GMM_{K}(dw|\pi,\mu+\offset(\theta),\Sigma)-\GMM_{K}(dw|\hat\theta)\big\rbrace,\\
		p_{\hat w}(d\hat w) &:= \min\big\lbrace\GMM_{K}(d\hat w|\hat\theta)-\GMM_{K}(d\hat w|\pi,\mu+\offset(\theta),\Sigma),0\big\rbrace.
	\end{align*}
	Since the distributions are disjoint and are identically mapped onto $\X$,
	we have $\psi(\Rx{})=0$ and it follows from Eq.~\eqref{eq:fullcoup_temp} that $\fWt(\Rx{})=\Wt(\Rx{})$.
\end{proof}

Analogous to the proof of \cite[Thm.~4.3]{lavaei2021compositional}, there exists a probability kernel $\fWt=\prod_{i=1}^{N}\fWt^i$ over $(\Xh\times\X,\mathcal{B}(\Xh\times\X))$ that couples $(\Trh,\Tr)$ over $\Rx{}$. 
Using Lem.~\ref{lem:maxcoup}, we obtain that $\fWt(\Rx{})=\prod_{i=1}^{N}\Wt^i(\Rx{i})$. Hence, following \cite[Thm.~1]{Schoen2023BayesSSR} we have that $\Wt:=\prod_{i=1}^{N}\Wt^i$ defines a sub-probability coupling of $(\Trh,\Tr)$ over $\Rx{}$.
Note that $\Wt$ satisfies all conditions for a valid sub-probability coupling of $(\Trh,\Tr)$ over $\Rx{}$:
\begin{align*}\SwapAboveDisplaySkip
	\Wt(\Xh\times\X) &= \prod_{i=1}^{N}\Wt^i(\Xh^i\times\X^i)= \prod_{i=1}^{N}\Wt^i(\Rx{i})= \Wt(\Rx{}),\\
	\Wt(d\hat x\times\X) &= \prod_{i=1}^{N}\Wt^i(d\hat x^i\times\X^i)\leq \prod_{i=1}^{N}\Trh^i(d\hat x^i) = \Trh(d\hat x),\\
	\Wt(\Xh\times dx) &= \prod_{i=1}^{N}\Wt^i(\Xh^i\times dx^i)\leq \prod_{i=1}^{N}\Tr^i(dx^i) = \Tr(dx).
\end{align*}
Since $\Wt(\Rx{}) \geq \prod_{i=1}^{N}(1-\delta^i)$, we have $\delta=1-\prod_{i=1}^{N}(1-\delta^i)$.
This concludes the proof of Thm.~\ref{thm:comp_induction_two}.
\end{proof}
	
\subsection{Proof of Thm.~\ref{thm:globalGuaranteesCascaded}}
\begin{proof}
	Due to the acyclic network structure, we can write the satisfaction in terms of conditional probabilities similar to Bayesian networks, namely
	\begin{align}
	\SwapAboveDisplaySkip
		&\P\left(\Ca\times\M\satisfies\psi\right) = \P\Big(\bigwedge\nolimits_{i=1}^N(\Ca^i\times\M^i\satisfies\psi^i)\Big),\label{eq:temp}\\
		&\hspace{15pt}= \prod_{i=1}^{N}\P\big(\Ca^i\times\M^i\satisfies\psi^i \given \Ca^j\times\M^j\satisfies\psi^j,\,\forall j\in\mathrm{Pre}(i) \big).\nonumber
	\end{align}
	Recall that a system $\Ca^j\times\M^j$ satisfies its specification $\psi^j$ if all its output traces $\mathbf{y}^j$ satisfy $\mathcal{L}^j(\mathbf{y}^j)\satisfies\psi^j$. We obtain \eqref{eq:globalGuaranteesCascaded} by taking the minimum of the terms in the right-hand side of \eqref{eq:temp} with respect to the trajectories in the condition set $\Ca^j\times\M^j\satisfies\psi^j$.
This concludes the proof.
\end{proof}
\subsection{Proof of Thm.~\ref{thm:globalGuaranteesCircular}}
\begin{proof}
	We add the safety specifications $\mathcal{A}^i$ to get
	\begin{align*}
		\P\left(\Ca\times\M\satisfies\psi\right) & \geq \P\left(\Ca\times\M\satisfies\psi\wedge \left(\wedge_{i = 1}^N \mathcal{A}^i\right)\right),\\
		& = 
		 \P\Big(\bigwedge\nolimits_{i=1}^N(\Ca^i\times\M^i\satisfies\psi^i)\andltl\mathcal{A}^i\Big).
	\end{align*}
	Next, we use a similar technique as in \cite[Thm.~3.3]{lavaei2022compositional} to get a lower bound based on the local satisfaction probabilities by considering the worst-case output traces of the individual predecessors $j\in\mathrm{Pre}(i)$ satisfying their local safety specification $\mathcal{A}^j$, i.e., $\mathbf{Y}^j:=\big\lbrace\mathbf{y}^j:\mathcal{L}^j(\mathbf{y}^j)\satisfies\mathcal{A}^j\big\rbrace$, and obtain \eqref{eq:globalGuaranteesCircular}.
\end{proof}

\subsection{Proof of Thm.~\ref{thm:gmmDelta}}
\begin{proof}
	The proof is similar to prior works \cite{Schoen2023BayesSSR, Schoen2022ParamUncert}, and showing that $\Wt$ in \eqref{eq:subprobcoup_subsys} is a sub-probability coupling of the systems \eqref{eq:subsystem} and \eqref{eq:sysNom} over relation \mbox{$\Rx{}$} in \eqref{eq:relation_subsys} is straightforward. 
	To compute the corresponding parameter $\delta$ we integrate $\Wt$ over the identity relation \mbox{$\Rx{}$}
	and relax using the trivial inequality $\min\big\lbrace \sum_i f_i, \sum_i g_i \big\rbrace \geq \sum_i \min\left\lbrace f_i, g_i \right\rbrace$ to get
	\begin{align*}
		&\Wt(\R|\theta) \geq
		\sum_{k=1}^{K}	\int_{\mathbb{W}}\int_{\X}
		\delta_{f(\hat\theta)+ w+\offset(\theta)}(dx) \delta_{f(\theta)+ w}(dx)\\
		&\hspace{20pt}\times
		\min\big\lbrace \pi_k \N(dw|\mu_k,\Sigma_k), \hat\pi_k \N(dw|\hat\mu_k-\offset(\theta),\Sigma_k)\big\rbrace,
	\end{align*}
where we omit the dependence on $(x,u)$ for conciseness.
	To reach a simpler formulation, we expand using the Cholesky decomposition $\Sigma_k=L_kL_k\T$, where $L_k$ is the lower triangular matrix of $\Sigma_k$.
	We rearrange and get
	\begin{align*}
		&\Wt(\R|\theta) \geq
		\sum_{k=1}^{K} \int_{\mathbb{W}}
		\min\big\lbrace \pi_k \N(dw|\beta_k,I),
		\hat\pi_k \N(dw|0,I)\big\rbrace,
	\end{align*}
	with $\beta_k:=\mu_k-\hat\mu_k+\offset$ and utilizing the
	choice of offset $\offset(x,u,\theta) = f(x,u;\theta)-f(x,u;\hat\theta)$ for all $(x,u,\theta)$.
	By similar reasoning as in \cite{huijgevoort2020similarityquantification} we split the individual integrals into integrations over disjoint halfspaces $E_k,\hat E_k$:
	\begin{equation*}
		\Wt(\R|\theta) \geq
		\sum_{k=1}^{K} \left[ \pi_k \int_{w\in E_k}\!\!\!\!\!\! \N(dw|\beta_k,I)
		+ \hat\pi_k \int_{w\in \hat E_k}\!\!\!\!\!\! \N(dw|0,I) \right]\!\!.
	\end{equation*}
	For each component, the halfspaces are given by
	\begin{equation*}
		\hat E_k:\quad \beta_k\T w > \eta_k\beta_k\T\beta_k,\quad
		E_k:\quad \beta_k\T w \leq \eta_k\beta_k\T\beta_k,
	\end{equation*}
	with $\eta_k:=\dfrac{1}{2} - \dfrac{1}{\beta_k\T\beta_k}\log\dfrac{\pi_k}{\hat\pi_k}$.
	We integrate and get
	\begin{align*}
		&\Wt(\R|\theta) \geq
		\sum_{k=1}^{K}  \pi_k \int_{w:\, \beta_k\T w \geq \eta_k\beta_k\T\beta_k} \hspace{-55pt} \N(dw|0,I)
		+ \hat\pi_k \int_{w:\, \beta_k\T w < \eta_k\beta_k\T\beta_k} \hspace{-55pt} \N(dw|\beta_k,I) ,\\
		&= \sum_{k=1}^{K}  \pi_k \int_{w:\, \beta_k\T w \leq -\eta_k\beta_k\T\beta_k} \hspace{-55pt} \N(dw|0,I)
		+ \hat\pi_k \int_{\hat w:\, \beta_k\T \hat w < (\eta_k-1)\beta_k\T\beta_k} \hspace{-55pt} \N(d\hat w|0,I) ,\\
		&= \sum_{k=1}^{K}  \pi_k \mathrm{cdf}\big(-\eta_k\norm{\beta_k}\big)
		+ \hat\pi_k \mathrm{cdf}\big((\eta_k-1)\norm{\beta_k}\big) ,
	\end{align*}
	where we use the invertability of the normal centered at zero in the first step.
	Hence, by considering the supremum over all $\theta\in\Theta$, we obtain Eq.~\ref{eq:deltafcn}.
\end{proof}

\end{document}